\documentclass[aip,jap,reprint]{revtex4-1}  

\usepackage{graphicx}
\graphicspath{{Figures/}}
\DeclareGraphicsExtensions{.eps,.ps}

\newcommand\txt[1]{\texttt{#1}}

\begin{document}

\title{A nanoscale transistor based on gate-induced stochastic transitions}

\author{J.~B\"urki}
\affiliation
  {Department of Physics and Astronomy, Sacramento State University, 
  6000 J Street, Sacramento, CA 95819}
\email{Buerki@csus.edu}

\author{C.~A.~Stafford}
\affiliation
  {Department of Physics,
  University of Arizona, 1118 E.\ Fourth Street, Tucson, AZ 85721}

\author{D.~L.~Stein}
\affiliation
  {Department of Physics and
  Courant Institute of Mathematical Sciences, New York University,
  New York, NY 10003}
\date{September 2010}
%


\begin{abstract}
  A nanoscale device consisting of a metal nanowire, a dielectric, and a 
  gate is proposed.  A combination of quantum and thermal stochastic effects 
  enable the device to have multiple functionalities, serving alternately as 
  a transistor, a variable resistor, or a simple resistive element with 
  $I-V$ characteristics that can switch between ohmic and non-ohmic. By 
  manipulating the gate voltage, stochastic transitions between different 
  conducting states of the nanowire can be induced, with a switching time as 
  short as picoseconds. With an appropriate choice of dielectric, the 
  transconductance of the device can significantly exceed the conductance 
  quantum $G_0=2e^2/h$, a remarkable figure of merit for a device at this 
  lengthscale.
\end{abstract}


\maketitle

\section{Introduction}
\label{sec:intro}
Variable resistors are commonly used circuit elements in many electronic
applications.  However, their large size and slow response time have
heretofore limited their use primarily to the human-circuit interface.
In this article, we describe how the exploitation of quantum and
stochastic effects at the nanoscale \cite{Burki05} allows one to
combine what would ordinarily be distinct macroscale circuit elements into
a single nanoscale device with multiple functionalities, and to achieve
response times on the order of picoseconds.

\begin{figure}[b]
  \centering
  \includegraphics[width=8.5cm]{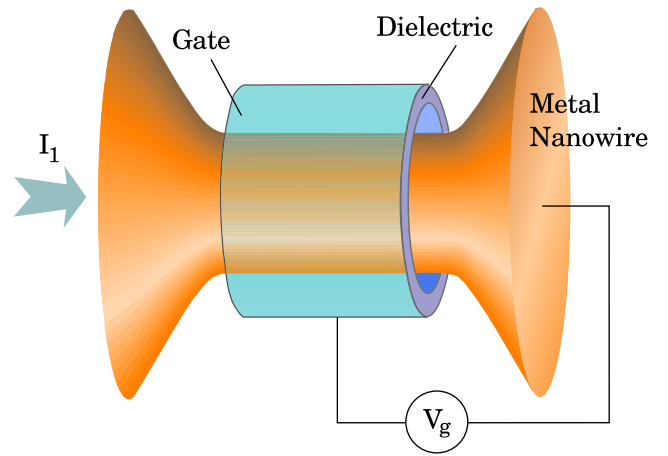}
  \caption{Diagram of the proposed device}
  \label{fig:diagram}
\end{figure}

The device architecture we propose is illustrated in Fig.~\ref{fig:diagram}.
The physics behind its operation is the following: A metal nanowire is the active circuit 
element, and is embedded in a dielectric sheath, surrounded by an outer 
conductor of comparable dimensions, referred to as the gate.  
A positive/negative voltage applied to the gate enhances/depletes the density 
of carriers in the nanowire.  
This results in a shift of the electronic Fermi energy $E_F$, which alters the 
electron-shell structure of the nanowire \cite{Urban03,Urban04,Urban06,Burki05}.
This in turn changes the energy barriers that determine the rates of stochastic 
transitions \cite{MS01,BSS05} between different conducting states of the nanowire.
The idea is that a sufficient voltage applied at the gate will make the energy 
barrier small enough to be comparable to the thermal energy $k T$, thus making 
a stochastic transition all but certain.

Such a device would require as active element a nanowire where electron-shell effects 
dominate over atomic-shell effects \cite{Mares04}. This has been shown \cite{Urban04b} 
to be the case for wires of conductance between a few and about a hundred conductance 
quanta $G_0=2e^2/h$.  
Such nanowires with lengths below or around a few nanometers have been fabricated 
using various techniques, including scanning tunneling microscopy (STM)
\cite{Agrait93,Rubio96}, mechanically-controllable break junctions (MCBJ) \cite{Yanson98,Yanson99},
thin-film transmission electron microscopy (TEM) \cite{Kondo97,Kondo00},
electromigration \cite{Strachan05}, and electrochemical fabrication
\cite{Tao02c}.
Nanowires with diameters less than a nanometer have been directly 
observed \cite{Kondo97} using TEM to remain stable under low beam 
intensities below $5\,\txt{A}/\txt{cm}^2$ for the duration of observation.

Stochastic switching between different conductance values has been observed 
in contacts made using MCBJ \cite{Krans96}, while controllable switching 
has been achieved recently using electromigration to grow or shrink 
a nanobridge between two wires \cite{Terabe05}.  
A structural thinning process of the nanowire similar to the one described 
by the theory of \citeauthor{BSS05} \cite{BSS05,Burki07} has been observed 
to take place for gold nanowires in TEM experiments.  
The nanowire was observed \cite{Oshima03} to thin step by step via a process
where a structural step (corresponding to a change in radius of the order 
one atomic diameter) forms at one end of the wire and subsequently 
propagates along the wire.  

A nanoelectromechanical switch similar to the device we are proposing has recently 
been built by \citealt{Martin09} on a somewhat larger scale.\nocite{Martin09}  
However, the larger scale of their device results in a 
different physical process responsible for the switching.
\citeauthor{Martin09} argue that the switching in their device is 
the result of a bending of the wire due to an electrostatic attraction between 
the wire and the gate.  
Such bending would be negligible in the device we envision, which would be much shorter 
(few nm as opposed to close to a $\mu $m in \citetext{Ref.\ \citenum{Martin09}}), and the switching would be instead 
driven by stochastic events as described above. 
The different scales, configurations, and physics underlying the operation of 
the two leads to different device characteristics and functionalities; 
these will be reviewed at the end of this article. 

\section{Background}
\label{sec:background}

Metal nanowires have attracted considerable interest in the past decade due
to their remarkable transport and structural properties \cite{Agrait03}.
Long gold and silver nanowires were observed to form spontaneously under
electron irradiation \cite{Kondo97,Rodrigues02b,Oshima03}, and appear to be
surprisingly stable.  Even the thinnest gold wires, essentially chains of
atoms, have lifetimes of the order of seconds at room temperature
\cite{Smit03a}.  
Metal nanowires exhibit striking correlations between their stability and
electrical conductance \cite{Urban04b,Mares07}.  
That these filamentary structures are stable at all is rather counterintuitive
\cite{Kassubek01,Zhang03}, but can be explained by electron-shell effects
\cite{Yanson99,Yanson01,Kassubek01,Zhang03,Burki03}.  

Because most of their atoms are at the surface, with low coordination numbers,
metal nanowires behave essentially like fluids~\cite{Zhang03}.
Classically, the Rayleigh instability would break up any wire whose length
exceeds its circumference~\cite{Kassubek01}.  
Nevertheless, nanowires clearly violating the Rayleigh criterion have
been observed repeatedly.\cite{Kondo97,Rodrigues02b,Oshima03}  
The instability is suppressed through quantum effects, with stabilization 
occurring through the nanowire's electronic shell structure.  
A quantum linear stability analysis~\cite{Kassubek01,Urban03,Urban04,Urban06} 
showed the existence of ``islands of stability'' for discrete intervals of the radius~$R$. 
These correspond to conductance ``magic numbers'' that agree with those observed in
experiments.  For low enough temperatures, there remain finite regions
of~$R$ stable against long-wavelength perturbations.  Therefore, stable
wires exist only in the vicinity of certain ``magic radii'' and
consequently at quantized conductance values $G$ that are integer multiples
of the conductance quantum $G_0=2e^2/h$.

However, the linear stability analysis ignores large thermal fluctuations that
can lead to breakup of the wire.  Nanowire lifetimes are inferred from conductance
histograms, compiled by cycling a MCBJ 
thousands of times~\cite{Yanson99,Yanson00,Yanson01}.  These studies indicate that
conductance peaks disappear above fairly well-defined temperatures, suggesting 
a lifetime at that temperature that is comparable to a typical wire-elongation time 
in the experiment.
\citealt{BSS05} studied the lifetimes of these
nanowires using techniques developed in \citetext{Ref.~\citenum{MS01}}.  
By modeling thermal fluctuations through stochastic Ginzburg-Landau classical 
field theories, they constructed a self-consistent approach to the fluctuation-induced
``necking'' of nanowires that is in good agreement with
experiment~\cite{BSS05,BSS06}.  Their theory indicates that passivated
noble metal nanowires are sufficiently stable at room temperature to serve as
interconnects between nanoscale circuit elements.

Of particular interest for the applications considered in this proposal
is the nature of the barriers separating wires of different magic radii
(and hence quantized conductances).  These barriers can be surmounted in
several ways: among them are raising the temperature, applying strain,
shortening the wire, or changing the Fermi energy.  The first three are
discussed in \citetext{Refs.\ \citenum{BSS05} and \citenum{BSS06}}, but the last is new to 
this paper.  For the purposes of a new nanoscale device, the first two may be
unsatisfactory for various reasons having to do either with nonoptimal
operating conditions (temperature), or probable inability to implement
these controls on the nanoscale (strain).

\section{Theoretical model}
\label{sec:theory}

\begin{figure}[bt]
  \centering
  \includegraphics[width=8.5cm]{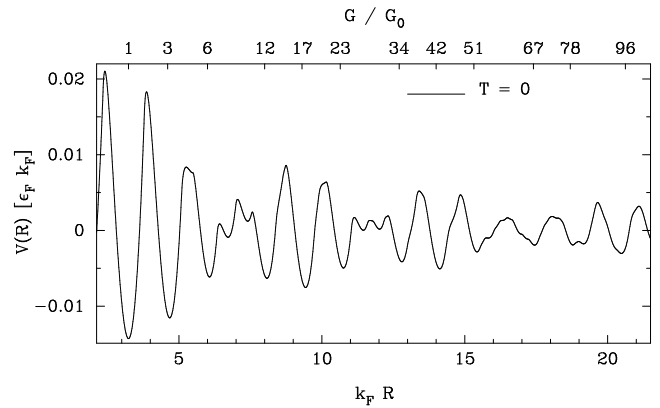}
  \caption{Electron-shell potential as a function of wire radius $k_F R$. The 
    conductance of the wires with ``magic radii'' is indicated on the top axis.
    From \citetext{Ref.\ \citenum{BSS05}} }.
  \label{fig:shell_potential}
\end{figure}

\begin{figure}
  \centering
  \includegraphics[width=8.5cm]{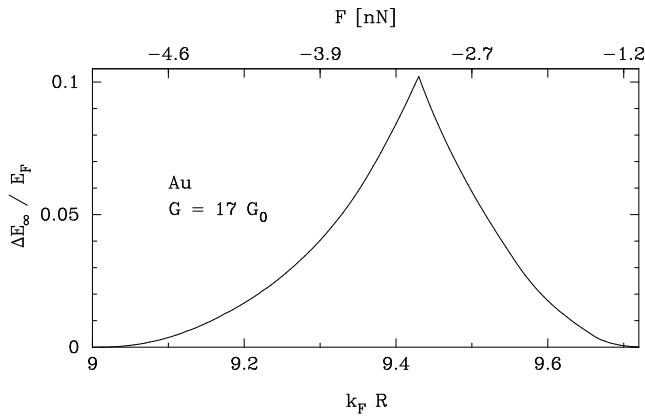}
  \caption{Escape barrier as a function of wire radius $k_FR$, or equivalently 
    applied stress $F$. Results correspond to a gold wire with a conductance 
    $G=17\,G_0$, following the calculations of \citetext{Ref.\ \citenum{BSS05}}. }
  \label{fig:strain}
\end{figure}

The magic radii are the minima of the electron-shell potential (see 
Fig.~\ref{fig:shell_potential}, and e.g.\ \citetext{Ref. \citenum{Burki03}}), which depends 
on the dimensionless parameter $k_F R$, with $k_F$ the Fermi wavevector 
and $R$ the wire radius.  
A shift in $k_F$ is thus analogous to applying strain, and can be used to
induce rapid (i.e., on the scale of the Debye frequency) transitions
between neighboring magic radii.  These have conductances differing by $n
G_0$, where $n\geq 2$ is an integer (see Fig.~\ref{fig:shell_potential}).  (As a
rule of thumb, the jumps scale as $n\sim (\pi/4)k_F R$ for a wire with
initial radius $R$.)  The switching time between two adjacent magic radii
was shown~\cite{BSS05} to be given by the Kramers formula
\begin{equation}
\label{eq:Kramers}
\tau \sim \Gamma_0^{-1} \exp(\Delta E/k_B T),
\end{equation}
where $\Delta E$ is the energy barrier, $k_B$ is Boltzmann's constant, and
$T$ is the temperature.  The rate prefactor $\Gamma_0$, of order the Debye
frequency, was calculated explicitly in \citetext{Ref.~\citenum{BSS05}}.  
The dependence of $\Delta E$ on the parameter $k_F R$ is illustrated in 
Fig.~\ref{fig:strain}.

The possibility of shifting $E_F$ electrostatically, as described above,
depends in an essential way on the crucial feature that the nanowire has a
radius of order nanometers, and thus has a very low density of states at
$E_F$.  As a function of the applied gate voltage~$V_g$, the shift
in $E_F$ is given by~\cite{Kassubek99}
\begin{equation}
\delta E_F = \frac{e V_g}{1 + (e^2/C_g)dN/dE}\, ,
\label{eq:dE_F}
\end{equation}
where $C_g$ is the mutual capacitance between gate and nanowire, and
$dN/dE$ is the density of states of the nanowire at $E_F$.  As discussed
in \citetext{Ref.~\citenum{Kassubek99}}, the denominator in Eq.~\ref{eq:dE_F} 
can be well approximated
in terms of material and geometrical parameters, yielding a convenient
rule-of-thumb estimate
\begin{equation}
\delta E_F \approx \frac{e V_g}{1 + \alpha \, r_s \, \epsilon^{-1}G/G_0},
\label{eq:dE_F_est}
\end{equation}
where $r_s$ is the Fermi gas parameter for the nanowire material
(essentially the mean inter-electron separation in the bulk metal), 
$\epsilon$ is the mean dielectric constant of the dielectric sheath,
and $\alpha$ is a dimensionless parameter of order unity, which depends
logarithmically on the device dimensions. 

In order to achieve the maximum switching speed, it is necessary to achieve
a shift $\delta (k_F R)\sim 1$ in the shell-potential parameter.  From
Eq.~\ref{eq:dE_F_est}, this implies a preferred operating gate voltage
\begin{equation}
\frac{e V_g}{E_F} \sim \frac{\alpha \, r_s \, k_F R}{6 \, \epsilon}.
\label{eq:dVg_est}
\end{equation}
For typical metals, $r_s \sim 2$--3, while $k_F R \sim 10$ in the domain of
validity of the nanoscale free electron model~\cite{Stafford97,Burki05}.  
It is therefore desirable to use a dielectric with~$\epsilon \geq 10$ to 
minimize the necessary gate voltages.

\subsection{Transconductance}
\label{sec:transconductance}

Because the mechanical switching time of the nanoscale variable resistor
can be as short as picoseconds, it may also be thought of as an~{\em
electromechanical transistor\/}.  It is thus useful to compute its~{\em
transconductance}, a figure of merit used to characterize transistors.  The
transconductance $g_T$ can be estimated as
\begin{equation}
\label{eq:transconductance}
g_T=\frac{dI_1}{dV_g} \sim \frac{n \, G_0 V_{12} }{V_g},
\end{equation}
where $V_{12}$ is the device bias voltage.  
Using Eq.~\ref{eq:dVg_est} and $n\sim (\pi/4)k_F R$, one finds
\begin{equation}
\label{eq:trans2}
\frac{g_T}{G_0} \sim \frac{3\pi \epsilon}{2\alpha r_s} \frac{e V_{12}}{E_F}.
\end{equation}
For large dielectric constants~$\epsilon\geq 10$, and bias voltages~$V_{12}
\sim 1$V, one can thus achieve $g_T\gg G_0$, an exceptional figure of merit
for a nanoscale device~\cite{CSM06}, thereby enabling its advantageous use 
as an effective transistor.  
In addition to the structural switching time of order picoseconds, the 
electrical $RC$ rise time $\tau=C_g/G$ can be estimated to be of order 
1 femtosecond for typical device dimensions/materials, and so
is not a limiting factor in device performance.

\begin{figure}[hbt]
  \centering
  \includegraphics[width=8.5cm]{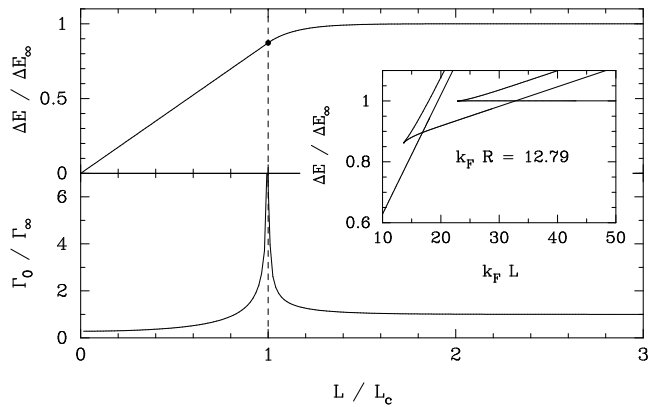}
  \caption{Escape barrier $\Delta E$ (top) and prefactor $\Gamma$ (bottom) 
    as a function of wire length, as calculated in \citetext{Ref.~\citenum{BSS05}.}  
    A second-order phase transition takes place 
    at the critical length $L_c$. For some wires, the transition is first-order 
    (inset).}
  \label{fig:transition}
\end{figure}
\subsection{Ohmic$\leftrightarrow$non-Ohmic transition}
\label{subsec:ohmic}

The device discussed above is one where barriers are controlled by shifting
the Fermi energy of the nanowire through electrostatic means.  Another
possibility is to change the wire length.  \citetext{In
Ref.~\citenum{BSS05}} it was predicted that a transition in activation
behavior occurs as a function of wire length: below a critical length
$L_c$, the barrier decreases rapidly with length, while above it is roughly
constant.  The transition can be continuous (second-order) or discontinuous
(first-order) (see Fig.~\ref{fig:transition}).  This effect may have already
been observed: a recent study~\cite{YOT05} reported a transition from
linear to nonlinear $I-V$ behavior in gold nanowires, as distance between
electrodes shortened due to an applied bias (presumably because of thermal 
expansion of the electrodes).  In a Comment~\cite{BSS06}, we
were able to explain this change in $I-V$ behavior as a consequence of the 
transition in radius stability as a function of wire length (cf.~Fig.~\ref{fig:transition}).  

\begin{figure}
  \centering
  \includegraphics[width=8.5cm]{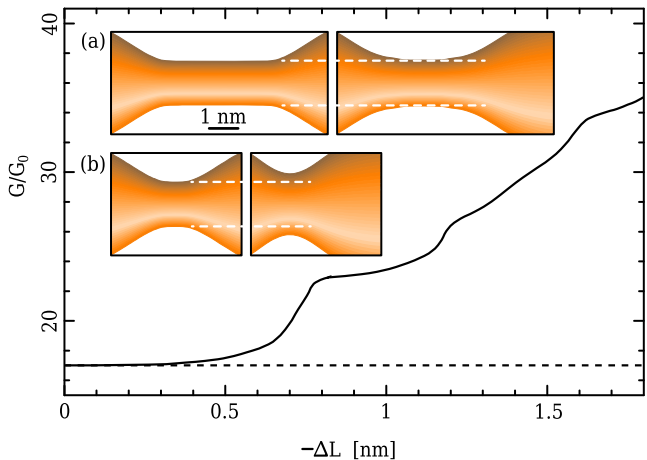}
  \caption{Conductance of a short (solid line) and long (dashed line) wire 
    under compression.  The initial and final shapes of the long (a) and 
    short (b) wires are shown in the inset.  From \citetext{Ref.~\citenum{BSS06}.}}
  \label{fig:IV}
\end{figure}

This suggests another device possibility, namely one where the wire length can be 
shortened below its critical value directly by increasing the applied voltage.  
This would convert a wire with linear $I-V$ characteristics to
one with nonlinear ones (see Fig.~\ref{fig:IV}).  At the present time, it
remains unclear how easily controllable such a transition might be and whether 
it would be reversible.

\section{Device realizations}
\label{sec:device}

Construction of a nanoscale variable resistor/electromechanical transistor 
in the laboratory will require combining the three components of the device---metal nanowire, 
dielectric, and gate--- in a single nanostructure.

\begin{figure}
  \centering
  \includegraphics[width=8.5cm]{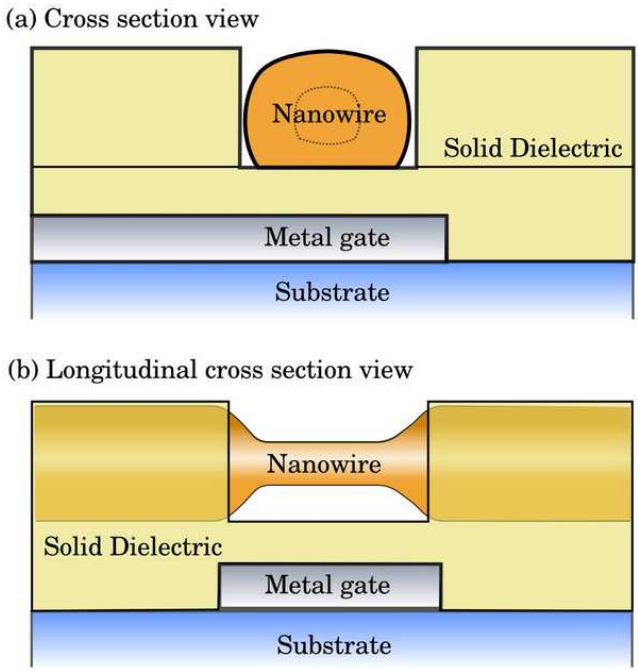}
  \caption{(a) Cross-section and (b) longitudinal cross-section diagrams of a device 
    with a solid dielectric, discussed in Example 2.  
    The gap shown around the nanowire may be filled with a liquid or gel dielectric 
    to improve performance.  
    Additionally, the cavity may be hermetically sealed, and a top gate can be added.}
  \label{fig:pit}
\end{figure}

To achieve optimal device characteristics, the space between the active segment 
of the nanowire and the gate(s) should be filled with a dielectric with 
$\epsilon \geq 10$.  
If a solid dielectric (only) is used, a small gap around the active segment of 
the nanowire must be provided (see Fig.~\ref{fig:pit}(b)) to permit the 
nanowire surface to fluctuate freely.  In that case, the mean dielectric constant of the
region between the nanowire and the gate(s) (including the gap) should exceed ten.  Many
intrinsic semiconductors could serve as suitable solid dielectrics with
$\epsilon \geq 10$ (e.g., Si, Ge, InSb, InAs, InP, GaSb, or GaAs).  The material should be
chosen so that the semiconducting energy gap exceeds the maximum desired voltage difference
between the gate and nanowire.

A liquid dielectric or combination of solid
and liquid dielectrics could also be utilized.
This would allow for optimal filling of the dielectric region, while still permitting free
motion of the nanowire surface.  
Liquid dielectrics have been used in conjunction with some of the previous
techniques \cite{Tao02a,Tao02b,Tao02c}, in the context of single
molecule measurements, as well as for STM measurements of metal contact
transport. In the latter context, they have been shown to have little influence
on the stability and transport properties of the nanocontact \cite{Tao02b}.

\begin{figure}[b]
  \centering
  \includegraphics[width=8.5cm]{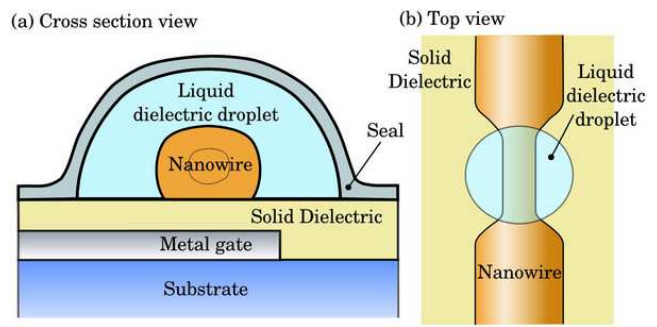}
  \caption{(a) Cross-section view and (b) top view diagrams of a device 
    with a nanodrop of liquid dielectric, discussed in Example 1. }
  \label{fig:droplet}
\end{figure}

Below we propose two architectures to realize the device described above.

\subsection*{Realization 1}

In a first possible architecture for the proposed device (see Fig.~\ref{fig:droplet}),
a layer of solid dielectric is deposited on
a substrate prepatterned (using standard vapor deposition techniques)
with a metallic gate to address the nanowire device.  A metal
nanowire several tens of nanometers in diameter with a ``notch'' or constriction at the
desired location
is then deposited on the surface of the dielectric, in alignment
with the submerged gate electrode.
The nanowire at the notch can then be thinned down to the specified operating diameter
by electromigration, scanning-electron microscopy (SEM), or chemical etching.
A nanodroplet of liquid dielectric is then deposited on the surface of the wafer,
immersing the nanowire device (see Fig.~\ref{fig:droplet}).
The nanowire device, together with the
droplet of liquid dielectric, is then hermetically sealed, e.g., with an epoxy seal.

\subsection*{Realization 2}

In another realization of the proposed device (see Fig.~\ref{fig:pit}),
a layer of solid dielectric is deposited on a substrate prepatterned with a 
metallic gate to address the nanowire device, as in the first realization.  A metal
nanowire of uniform diameter several tens of nanometers
is then deposited on the surface of the dielectric, in alignment with the 
submerged gate electrode.  This fabrication step can be carried out within standard 
semiconductor patterning techniques, such as for example E-beam direct write or 
alternatively in the long term masked ion beam lithography. 
A further layer of solid dielectric is then deposited, fully encasing the nanowire.  
A nanoscale pit or cavity is then etched in the dielectric layer,
exposing the active segment of the nanowire.  The exposed segment of the nanowire 
is then thinned down to the specified diameter (of order one nanometer) via e.g., 
focused SEM, chemical etching, or electromigration, or a combination of these techniques. 

A hermetic seal can be applied to increase the durability of the nanowire device.
For example, an epoxy bubble seal may be used to enclose an inert
atmosphere (e.g., nitrogen or argon) about the exposed segment of the nanowire.  
Alternatively, a passivation layer over
the nanowire device is used to scavenge any small amounts of oxidant from 
the sealed environment. 

The pit containing the nanowire (see Fig.~\ref{fig:pit}(b)) can also be filled 
with a liquid or gel dielectric before the seal is applied, to enhance 
device performance.

For some applications, both a top gate (not shown) and a bottom gate are included, 
above and below the nanowire device, respectively.  
Multiple gates may be desirable e.g.\ to address individual devices in a large array.  
For example, if the gate voltage is chosen appropriately, the device will switch 
conducting states rapidly only if the voltage is applied to both gates.

Because the three terminals of the nanoscale variable resistor consist of metal 
which can be patterned by standard semiconductor fabrication techniques, 
such devices can be readily integrated with conventional circuitry on a chip.  
Because the throughput impedance of such a device is on the scale of several 
hundred to several thousand Ohms, appropriate amplification may be required 
to interface with standard CMOS circuitry.

\section{Discussion}

We have proposed a nanoscale device that exploits the new physics that
emerges at the boundary between classical and quantum physics.  A
monovalent metallic nanowire is stable at certain ``magic radii''
\cite{Burki03}, corresponding to a subset of integer multiples of the
quantum conductance $G_0=2e^2/h$.  The nanometer lengthscale of these
systems leads to low energy barriers between different conductances that
are surmountable through classical thermal fluctuations on short
timescales.  The barriers themselves are functions of both the wire's
material parameters --- for example, electron density of states --- and
wire geometry, and as such are controllable through external manipulation.
This leads to a novel situation where a basic wire configuration can
perform multiple functions that on the macroscopic scale require different
specially designed electronic components.

The configuration of the device --- a metal nanowire embedded in a
dielectric sheath surrounded by an outer conductor, which serves as the
gate --- is simple. Changing the gate voltage alters the density of
carriers, which in turn allows sensitive control of the energy barriers
that determine the rates of stochastic transitions between different
conducting states of the nanowire \cite{BSS05}.  The device can
alternatively serve as a simple resistive element, as a variable resistor,
or as an electromechanical transistor, given a switching time that can be
of the order of picoseconds.  A further possibility, though one that may be
more difficult to control, is using external voltage to change between
linear and nonlinear $I-V$ response, through a transition in activation
behaviors as noted in \citetext{Ref.\ \citenum{BSS06}}.

The device has some similarities with one built by \citealt{Martin09}, 
but there are important differences.  The latter, which
is realized on the $\mu$m rather than the nm lengthscale, depends on a
mechanical process (wire bending) for its operation, leading to a slower
response which effectively reduces the array of functionalities of the
device.  On the other hand, the device of \citetext{Ref.~\citenum{Martin09}} 
can serve as an on-off switch, whereas the device discussed here can change only 
between different nonzero conductances.  The two devices are therefore best viewed
as complementary, with each specialized to different but useful functions.

\acknowledgments

This work was supported by NSF Grant Nos.~0312028 (CAS), and PHY-0651077
and PHY-0965015 (DLS).  Part of this work was done when CAS and DLS were at
the Aspen Center for Physics, and JB was at the University of Arizona.  The
authors acknowledge the hospitality of the Santa Fe Institute where the
last stages of this work were completed.

%


\end{document}